\documentclass[prb,twocolumn,showpacs]{revtex4}

\usepackage{graphicx}

\begin{document}

\title{Magnetically induced spin-dependent photoemission from $p$-GaAs(Cs,O) into vacuum}

\author{D. A. Orlov}
\affiliation{Institute of Semiconductor Physics, Russian Academy
of Sciences, 630090, Novosibirsk, Russia,\\ and Max-Planck
Institut f$\ddot{u}$r Kernphysik, D-69117 Heidelberg, Germany}

\author{V. L. Alperovich}
\email{alper@thermo.isp.nsc.ru} \affiliation{Institute of
Semiconductor Physics, Russian Academy of Sciences, and
Novosibirsk State University, 630090, Novosibirsk, Russia}

\author{A. S. Terekhov}
\affiliation{Institute of Semiconductor Physics, Russian Academy of Sciences, and Novosibirsk State University,
630090, Novosibirsk, Russia}

\date{\today}

\begin{abstract}
A spin-dependent emission of optically oriented electrons from
$p$-GaAs(Cs,O) into vacuum was experimentally observed in a
magnetic field normal to the surface. This phenomenon is explained
within the model which takes into account the jump in the electron
$g$ factor at the semiconductor-vacuum interface. Due to this
jump, the effective electron affinity on the semiconductor surface
depends on the mutual direction of optically oriented electron
spins and the magnetic field, resulting in the spin-dependent
photoemission. It is demonstrated that the observed effect can be
used for the determination of spin diffusion length in
semiconductors.
\end{abstract}


\maketitle

\section{Introduction}

Investigations of spin-polarized electron transport through
various interfaces in solid-state structures constitute an
important domain of spintronics and have been an area of active
research for over ten years.\cite{Awschalom1} In this context,
studies of the emission process of optically oriented electrons
from semiconductors with the state of effective negative electron
affinity\cite{Bell2} (NEA) are of both applied and fundamental
interest. The application prospects are related to the development
of spin detectors for low energy electron
beams\cite{Cacho3,Alper4} and of efficient spin-polarized electron
sources.\cite{Pierce5,Prepost6} The scientific interest lies in
the elucidation of mechanisms of spin relaxation during transport
through a semiconductor-vacuum interface. It is also important
that this interface can be considered as a model one at which the
energy, effective mass and $g$ factor undergo jumps at ultimately
small distances of about an interatomic separation. In contrast to
the solid-state heterojunctions, in vacuum it is possible to
measure not only a total electric current through the interface
but also the energy and momentum distributions of emitted
electrons\cite{Drouhin7,Terek8,Orlov9} and their
spin.\cite{Drouhin10} These investigations enable one to clarify
the conditions and restrictions for the use of effective electron
parameters, such as the effective mass and the effective $g$
factor, for the description of charge and spin transfer through
abrupt interfaces. Measurements of photoemission with angular and
spin resolution open the possibility of studying spin-dependent
tunneling of electrons that was predicted in
Refs.~\onlinecite{Voskoboynikov11,Perel12,Tarasenko13}.

Spin-dependent electron transport, which is caused by the jump in
the electron $g$ factor at internal semiconductor interfaces, was
previously studied in Refs.~\onlinecite{Gruber14} and
\onlinecite{Fabian15}. Gruber et al.\cite{Gruber14} investigated
spin-dependent resonant electron tunneling through the Zeeman
levels of a double-barrier structure with a quantum well made of a
semimagnetic semiconductor. Fabian et al.\cite{Fabian15}
theoretically analyzed the dependence of the $I-V$ characteristics
of a magnetic $p-n$ junction on the direction of the electron spin
with respect to the magnetic field. The phenomena related to the
jump of the electron $g$ factor at a semiconductor-vacuum
interface have not yet been investigated. In this paper we study
the transport of spin-polarized electrons through the
$p$-GaAs(Cs,O)-vacuum interface. The probability of electrons
escaping into vacuum was found to depend on the direction of the
electron spin with respect to the magnetic field applied
perpendicularly to the surface. The experimental results are
described well in the emission model which takes into account the
jump in the electron $g$ factor at the semiconductor-vacuum
interface. The observation of the effect was preliminarily
reported in Ref.~\onlinecite{Orlov16}. In this paper we present
data and analysis which yield a self-consistent picture of the
spin-dependent photoemission at a semiconductor-vacuum interface.
The opportunity to explore this effect for studying spin transport
in semiconductors is demonstrated. In particular, the spectrum of
the observed effect allowed us to determine the spin diffusion
length of electrons in $p$-GaAs.

\section{Spin-dependent photoemission: a qualitative explanation}

\begin{figure}
\includegraphics{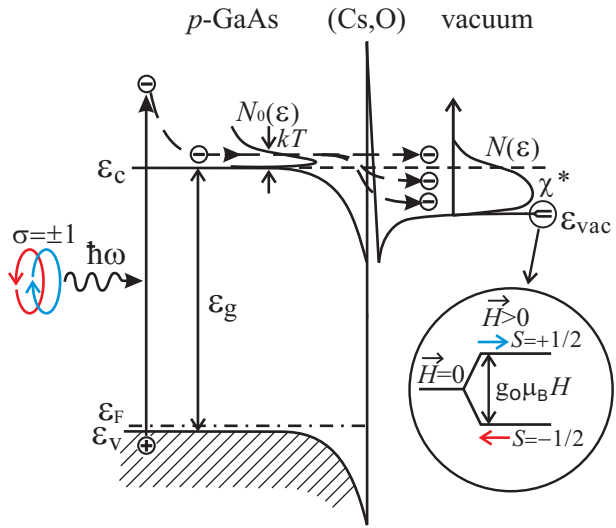}
\caption{(Color online) The mechanism of magnetically induced
spin-dependent photoemission illustrated on the energy band
diagram of NEA-photocathode. The generation of spin-polarized
photoelectrons by right ($\sigma=+1$) and left circularly
polarized light ($\sigma=-1$), their thermalization to the bottom
of the conduction band, diffusion toward the surface, and emission
into vacuum are schematically shown. $N_0(\varepsilon)$ and
$N(\varepsilon)$ are the energy distributions of thermalized
photoelectrons in the conduction band and of the electrons emitted
into vacuum, respectively. The Zeeman splitting of electron
states, caused by the external magnetic field, is shown
schematically for the vacuum level.}
\end{figure}
A phenomenological expression for the spin-dependent component
$J_S$ of photocurrent $J$ in a magnetic field $\textbf{H}$ can be
written in the following form:
\begin{equation}
J_S=C(\textbf{S}\cdot\textbf{H})J,
\end{equation}
where $\textbf{S}$ is the mean spin of optically oriented
electrons and the constant $C$ depends on the microscopic
mechanism of the effect and determines its relative magnitude.
Several microscopic mechanisms of magnetically induced
spin-dependent photoemission (SDP) are possible with both bulk and
surface origin. Preliminary estimations show that for the case of
spin-polarized electrons emitted from $p$-GaAs activated by cesium
and oxygen to the state of NEA, a mechanism based on the jump in
the $g$ factor at the semiconductor-vacuum interface can
significantly contribute to SDP. The jump in the $g$ factor causes
the difference between the NEA values for electrons with spins
oriented along or opposite to the magnetic field. This difference
results in the dependence of the photoemission current on the
direction of electron spin with respect to the magnetic field.
This mechanism is illustrated on the energy band diagram of the
semiconductor-vacuum interface (Fig.~1) and can be explained as
follows. The electrons, which are excited in the conduction band
by light with photon energies $\hbar\omega$ exceeding the band gap
$E_g$, are thermalized to the band bottom, form a narrow energy
distribution with the width of $\approx kT$, diffuse towards the
emitting surface, pass through the band bending region, and escape
into vacuum. Because of the momentum and energy scattering during
electron transport across the band bending region and through the
(Cs,O) activation layer\cite{Drouhin7,Terek8,Orlov9}, in vacuum
the kinetic energy distribution of electrons is broadened up to
the magnitude of NEA, $\chi^*$, which is defined as the energy
difference between the vacuum level and the bottom of the
conduction band in the bulk. The electrons, which descend along
the energy scale below the vacuum level during thermalization in
the band bending region, recombine at the surface and do not
contribute to the photoemission current. Therefore, the
photoemission quantum yield, as well as the photocurrent, depends
on the value of $\chi^*$.

The external magnetic field causes the Zeeman splitting of
electron states. At the bottom of GaAs conduction band the
effective $g$ factor is negative $g^*=-0.44$,\cite{Weisbuch17}
while in vacuum $g_0=2$. As a result, when the direction of the
electron spin with respect to the magnetic field changes, the
effective electron affinity $\chi^*$ changes by
$\Delta\chi^*=(g_0-g^*)\mu_BH$, where $\mu_B=e\hbar/2mc$ is the
Bohr magneton and $m$ is the free electron mass. As a result of
this change, when spin-polarized electrons are generated in the
conduction band by circularly polarized light\cite{Meier18}, the
photoemission current depends on the degree and sign of the
circular polarization.

\section{Experiment}

The experiments were carried out in a planar vacuum photodiode
consisted of a transmission-mode GaAs(Cs,O) photocathode bonded to
a glass substrate and a copper anode. The cathode and anode were
hermetically sealed parallel to each other on the opposite ends of
an alumina ceramic cylinder. The active $p$-GaAs layer of the
AlGaAs/GaAs heterostructure photocathode was doped by Zn up to
$p\approx5\times10^{18}$\,cm$^{-3}$. The photodiode was placed in
a solenoid with the magnetic field perpendicular to the
photocathode surface. In our experiments, magnetic fields up to
$H=1$\,Tesla were used. All measurements were performed at room
temperature. The photoemission current $J$ was measured by
illumination of the photocathode through the glass substrate by
light from a monochromator with a halogen lamp. The vacuum
photodiode can be used as an electron energy analyzer with a
uniform retarding electric field.\cite{Terek8} Specifically, by
varying the voltage $U$ applied between the anode and cathode it
is possible to collect on the anode all of the emitted electrons
(at $U>U_0$), or to collect only part of the emitted electrons
with kinetic energies $\varepsilon_{\parallel}>U_0-U$ (at
$U<U_0$). Here, $U_0$ is the voltage that is required to
compensate the work function difference and to establish zero
electric field in the space between the cathode and anode. The
longitudinal kinetic energy is defined as
$\varepsilon_{\parallel}=p_{\parallel}^2/2m$, where
$p_{\parallel}$ is the momentum component parallel to the total
photocurrent vector (and perpendicular to the surface).
\begin{figure}
\includegraphics{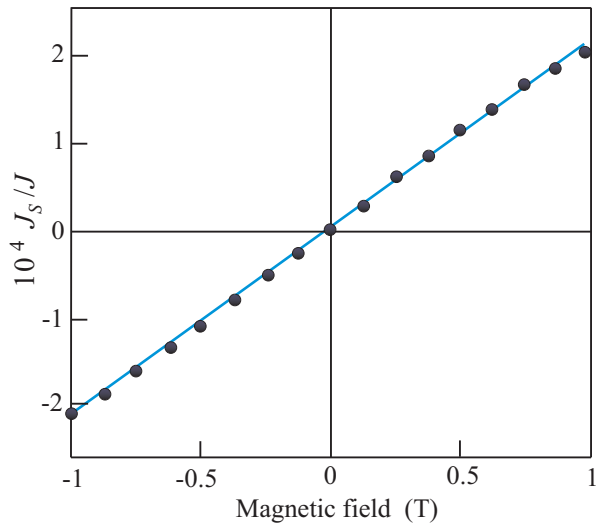}
\caption{(Color online) The magnetic field dependence of the
polarization-dependent component $J_S$ of the photocurrent
normalized to the total photoemission current $J$. The anode
voltage $U=+5$\,V, at which all emitted electrons are collected on
the anode. The photon energy $\hbar\omega=1.52$\,eV. The parasitic
component arising due to modulation of the light intensity is
subtracted (see text).}
\end{figure}

The spin-dependent component $J_S$ of the photocurrent was
measured by a lock-in amplifier as the difference
$J_S=J(\sigma^+)-J(\sigma^-)$, where $J(\sigma^+)$ and
$J(\sigma^-)$ are the photoemission currents for the excitation by
right and left circularly polarized light, respectively. The light
polarization was switched between the  $\sigma^+$ and $\sigma^-$
states with a frequency of 1.5 kHz by means of a wide-aperture
polarization modulator based on the linear electro-optic effect in
Bi$_{12}$SiO$_{20}$.\cite{Alper19} The modulation of the
polarization of the light beam was accompanied by a parasitic
modulation of its intensity with a relative value of about
$3\times10^{-4}$. In order to exclude the influence of the
intensity modulation on the results of the measurements, we took
into account that SDP is an odd function of the magnetic field. To
this end, the magnitude of $J_S$ was measured for two opposite
directions of the magnetic field $\pm$$H$, and the value of SDP
was determined as $\widetilde{J_S}= [J_S(+H)-J_S(-H)]/2$ (the
tilde over $J_S$ will be omitted in the text below). To justify
the possibility of using this procedure, the linearity of $J_S$ on
the magnetic field was tested. Figure 2 shows a typical magnetic
field dependence $J_S(H)$ measured at $\hbar\omega=1.52$\,eV. At
this photon energy, SDP caused by the jump in the $g$ factors
gives the major contribution to the measured effect (see Fig.\,3
and the text below). It is seen that in the investigated range of
magnetic fields, $J_S$ is a linear function of the magnetic field,
in accordance with phenomenological relation (1) and microscopic
mechanism of the effect described in Sec. II. In the text below,
basing on the linearity on $H$, we present experimental data
related to the maximal field $H=1$\,T, which yield the highest
signal to noise ratio.

\section{Results and discussion}

In addition to the spin-dependent photoemission caused by the jump
in the electron $g$ factor at the semiconductor-vacuum interface,
phenomenological equation (1) allows for the existence of bulk
effects such as magnetically induced circular dichroism
\cite{Seisyan20,Kaufmann21} and spin-dependent recombination
\cite{Solomon22,Weisbuch23,Paget24} in the bulk of GaAs, which may
also contribute to the experimentally measured values of $J_S$. In
GaAs, magnetically induced circular dichroism was previously
studied for the impurity-related optical transitions below the
band gap.\cite{Kaufmann21} To determine the magnitude of the
circular dichroism for the above band gap optical transitions, we
measured the relative change in the optical transmission of the
GaAs photocathode under the change in the sign of light circular
polarization $\Delta T/T=2
[T(\sigma^+)-T(\sigma^-)]/[T(\sigma^+)+T(\sigma^-)]$. Figure~3
shows the spectrum of $\Delta T/T$ measured in the magnetic field
$H=1$\,T, as well as the spectrum of $J_S/J$ measured at $H=1$\,T
and $U=+5$\,V, when all emitted electrons are collected on the
anode. From the comparison of the shapes of these two spectra, one
can assume that circular dichroism yields the main contribution to
SDP for photon energies near the band gap
$\hbar\omega\approx1.4$\,eV; however, this is not the fact for
$\hbar\omega>1.5$\,eV. It is also seen that for
$\hbar\omega=1.53$\,eV and 1.76\,eV, dichroism vanishes, while
$J_S/J$ is nonzero over the entire spectral range up to
$\hbar\omega=1.8$\,eV.
\begin{figure}
\includegraphics{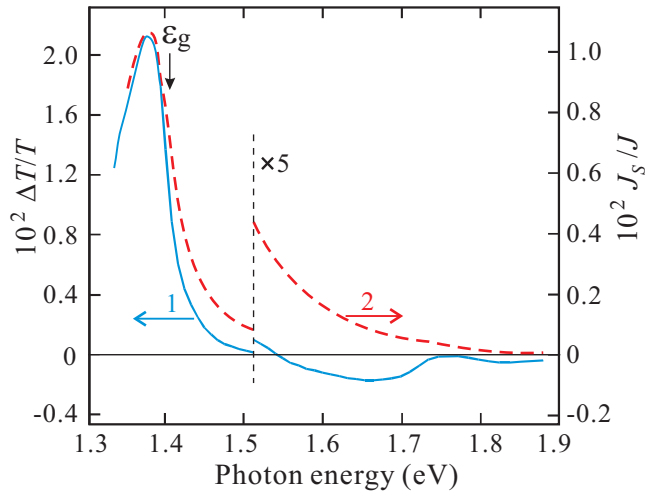}
\caption{(Color online) Spectra of circular dichroism $\Delta T/T$
(curve 1, left scale) and spin-dependent photoemission $J_S/J$
(curve 2, right scale). The magnetic field was equal to 1\,T.}
\end{figure}

It was not possible to determine the contribution of magnetically
induced spin-dependent recombination in the semiconductor bulk.
Therefore, in order to extract the surface contribution to the
spin-dependent photoemission, the procedure of measuring SDP was
modified so that all bulk contributions, including dichroism and
spin-dependent recombination in the bulk of GaAs, were subtracted
from the measured values of $J_S/J$. To this end, the magnitude of
$J_S/J$ was measured at various voltages $U$ between the anode and
cathode in the range of $U_0-\chi^*<U<U_0$. The lower limit
$(U_0-\chi^*)$ corresponds to an almost total cut-off of the
photocurrent, when only a small part of electrons emitted into
vacuum above the energy of the bottom of the conduction band in
the bulk can reach the anode. The upper limit $(U_0)$ corresponds
to a complete collection of all photoemitted electrons on the
anode. To determine the limits of this range
experimentally,\cite{Terek25} we measured the potential derivative
of the photoemission current $dJ/dU$, which is proportional to the
energy distribution function $N(\varepsilon_{\parallel})$ of
emitted electrons (solid curve in Fig.~4). As seen in Fig.~4, the
distribution function $N(\varepsilon_{\parallel})$ is bell shaped.
The width of $N(\varepsilon_{\parallel})$ is equal to the
magnitude of NEA;\cite{Terek8} in our case $\chi^*\approx
0.2$\,eV. The lower limit of the voltage range $U=2.7$\,V
corresponds to electrons which are emitted from the bottom of the
conduction band $\varepsilon_c$ without momentum and energy
scattering with a kinetic energy in vacuum
$\varepsilon_{\parallel}=\chi^*$. The upper limit $U=U_0=2.9$\,V
corresponds to electrons at the vacuum level $\varepsilon_{vac}$.
In a previous work,\cite{Terek25} we described in detail the
procedure for measuring the distribution function of electrons
$N(\varepsilon_{\parallel})$ as well as for the energy
calibration, which makes it possible to determine $U_0$ and to
interlink the voltage scale $U$ with the energy scale
$\varepsilon_{\parallel}$ and the position $\varepsilon_c$ of the
bottom of the conduction band in the semiconductor bulk.
\begin{figure}
\includegraphics{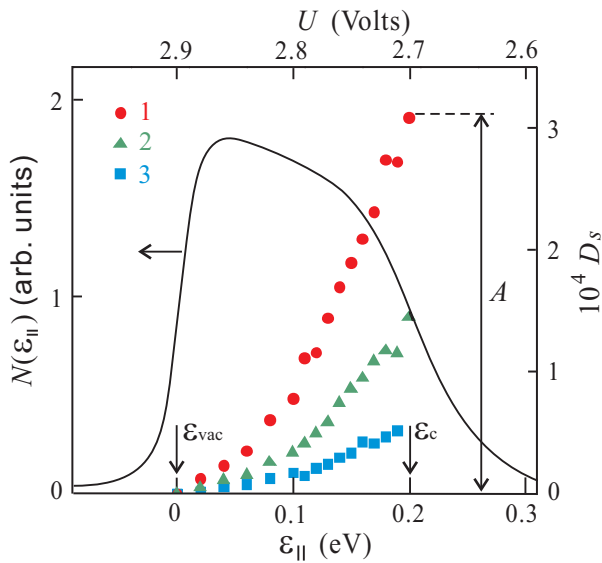}
\caption{(Color online) Energy distribution of emitted electrons
$N(\varepsilon_{\parallel})$ (solid curve, left scale) and voltage
dependences of the differential SDP $D_S(U)$ (data points, right
scale) measured at $H=1$\,T for various photon energies $\hbar\omega=1.42$\,eV
(1), 1.74\,eV (2), and 1.82\,eV (3). The energies
corresponding to the vacuum level $\varepsilon_{vac}$ and to the
bottom of the conduction band $\varepsilon_c$ are denoted by the
vertical arrows.}
\end{figure}

In order to extract the surface contribution to SDP, we calculated
the voltage-dependent part $D_S(U)=[J_S(U)/J(U)-J_S(U_0)/J(U_0)]$
of the spin-dependent photocurrent. By varying the retarding
voltage $U$, we measured the SDP for different groups of electrons
which undergo energy and momentum relaxation at the surface and
are emitted into vacuum with longitudinal energies below the
bottom of the conduction band in the bulk. For a zero surface
spin-dependent contribution, the value of $J_S/J$ must be the same
for different groups of scattered electrons. Therefore, the
voltage-dependent part of SDP can be assigned to only the surface
contribution because bulk contributions do not depend on $U$.

The data points in Fig.~4 show the dependence $D_S(U)$ measured at
various photon energies $\hbar\omega$. It is seen that at each
$\hbar\omega$ the differential spin-dependent photocurrent is
maximal for electrons emitted from the bottom of the conduction
band and monotonically goes down with increasing $U$, that is,
with decreasing electron kinetic energy. It is also seen from
Fig.~4 that the amplitude $A$ of variations in $D_S$ decreases
with increasing photon energy, while the shapes of the dependences
$D_S(U)$ are similar. Figure 5 shows the scaled dependences
$D_S(U)$ measured at various $\hbar\omega$. The scaling factors
were fitted to minimize the difference between the measured
dependences. It is seen that the shapes of all three dependences
$D_S(U)$ coincide with each other within the experimental
accuracy. Thus, the shape of the energy dependence $D_S(U)$ is
indeed independent of the photon energy.
\begin{figure}
\includegraphics{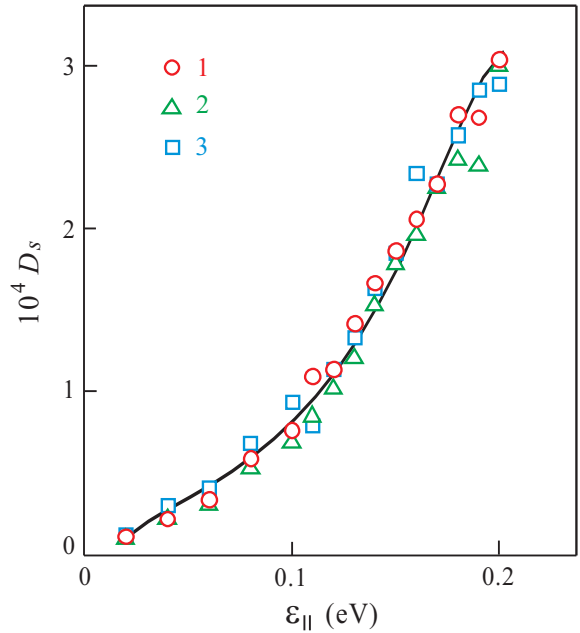}
\caption{(Color online) Differential SDP (data points) measured at
various photon energies $\hbar\omega=1.42$\,eV (1), 1.74\,eV (2),
and 1.82\,eV (3). Dependences (2) and (3) are multiplied by the
factors of 2.1 and 5.6, respectively, in order to scale with (1).
The solid line is the approximation of the measured dependences by
formula (2).}
\end{figure}

We compared the universal shape of the measured dependences
$D_S(U)$ to the calculation in the model of magnetically induced
spin-dependent photoemission arising due to the jump in the $g$
factor at the GaAs(Cs,O)-vacuum interface. In this model the
effective NEA and, consequently, the width of the energy
distribution function of spin-polarized electrons vary by
$\Delta\chi^*=(g_0-g^*)\mu_BH$ when the direction of the electron
spin changes with respect to the magnetic field. The respective
change in the current created by photoelectrons with mean spin $S$
is equal to $J_S=S\Delta\chi^*N(U)$. Thus, the voltage-dependent
component of SDP can be expressed as follows:
\begin{equation}
D_S(U)=S\Delta\chi^*\frac{N[\varepsilon(U)]}{J(U)}-D_0.
\end{equation}

This calculated dependence $D_S(U)$, which approximates the
experimental dependences, is shown by the solid line in Fig.~5.
For the calculation, we used the experimentally measured energy
distribution function $N(\varepsilon_{\parallel})$ and the voltage
dependence of the photoemission current $J(U)$ obtained by
numerically integrating this distribution. Taking into account a
finite resolution $\Delta\varepsilon_{\parallel}=20$ meV of the
measurements of the energy distribution function
$N(\varepsilon_{\parallel})$, Eq. (2) is valid for voltages
$U<U_0-\Delta\varepsilon_{\parallel}$. The value of the mean spin
of photoemitted electrons was determined in the diffusion
model\cite{Dzhioev26} from the spectral dependence of SDP
amplitude (see Fig.~6). The value of the electron affinity
modulation $\Delta\chi^*$, which determines the amplitude $A$ of
variations in SDP with varying $U$, was a fitting parameter and
found to be equal to 0.09\,meV. The constant $D_0\approx10^{-4}$
was determined so that the calculated value of $D_S$ coincides
with the experimental value for
$U=U_0-\Delta\varepsilon_{\parallel}$. It is seen that the shape
of the calculated dependence $D_S(U)$ describes the experiment
well. The fitting parameter $\Delta\chi^*=0.09$\,meV should be
compared with $\Delta\chi^*=0.14$\,meV estimated from the known
value $g^*=-0.44$ of the electron $g$ factor on the bottom of the
conduction band of GaAs. The agreement between the experimental
and calculated values of $\Delta\chi^*$ may be considered as
reasonably good, keeping in mind that photoelectrons acquire
kinetic energy in the band bending region before emission, and
this alters the effective $g$ factor due to its energy
dependence.\cite{Pfeffer27} Moreover, according to
Refs.~\onlinecite{Bell2,Korotkikh28,Orlov29}, in a semiconductor
with the state of NEA, the photoemission occurs via electron
capture to a two-dimensional sub-band in the band bending region
and subsequent elastic or inelastic tunneling into vacuum through
the potential barrier formed by the (Cs,O) layer. Therefore, the
effective $g$ factor may be changed by the electron
quantization,\cite{Ivchenko30,Kiselev31,Pfeffer32} as well as by
the atomic structure of the GaAs(Cs,O) interface. The magnitude
and energy dependence of SDP are also possibly influenced by the
partial relaxation of the spin of electrons passing through the
GaAs(Cs,O)-vacuum interface.\cite{Drouhin10} It is worth noting
that along with the jump in the $g$ factor, a microscopic cause of
the SDP at the surface may consist of spin-dependent recombination
of photoelectrons on paramagnetic surface centers oriented by the
magnetic field.\cite{Lepine33} However, spin-dependent
recombination likely did not play a significant role in our
experiment performed at room temperature.

The effect of spin-dependent photoemission can be used as a method
for determining the spin diffusion length $L_S$ of electrons in
semiconductors. The idea of this method can be explained as
follows. Under photocathode illumination the spatial distribution
of generated photoelectrons depends on the photon energy in
accordance with the spectral dependence of the light absorption
coefficient $\alpha(\hbar\omega)$. Therefore, the photons with
various energies $\hbar\omega$ generate electrons at various
distances from the emitting surface. Due to electron spin
relaxation in the course of the diffusion toward the emitting
surface, the mean spin of emitted electrons $S$ depends on
$\hbar\omega$. As the amplitude $A$ of the surface contribution to
SDP is proportional to $S$, this leads to the spectral dependence
of $A=A(\hbar\omega)$. The shape of this dependence is determined
by the spin diffusion length of electrons.
\begin{figure}
\includegraphics{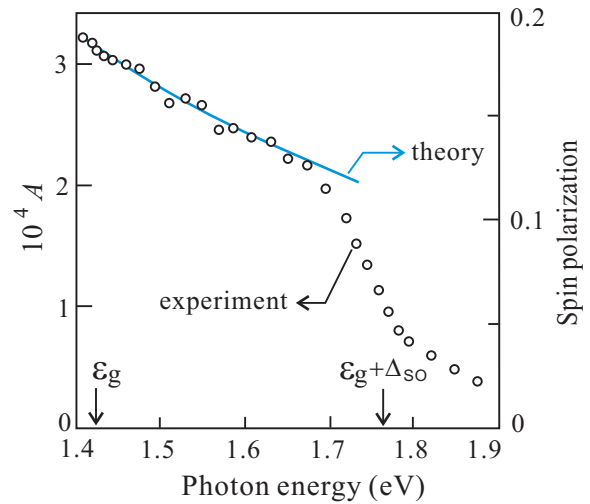}
\caption{(Color online) The spectra of the experimentally measured
amplitude $A$ of the surface contribution to SDP (data points,
left scale) and of the calculated degree of spin polarization of
photoemitted electrons (solid line, right scale). The amplitude of
SDP, $A$, was determined as $D_S(\varepsilon_c)$ (see Fig.~4).}
\end{figure}

The measured spectrum of $A(\hbar\omega)$ is shown in Fig.~6 by
dots (left scale). It is seen that the amplitude of SDP decreases
with increasing $\hbar\omega$. The abrupt drop in $A$ near
$\hbar\omega=\varepsilon_g+ \Delta_{so}\approx1.76$\,eV is due to
the onset of optical interband transitions from the spin-orbit
split valence band because these transitions generate electrons
with spin opposite to the mean spin of electrons excited by
near-band-gap optical transitions.\cite{Meier18} The gradual
decrease in the SDP amplitude $A$ at photon energies
$\varepsilon_g<\hbar\omega<\varepsilon_g+\Delta_{so}$ is due to
the dependence of the initial polarization of photogenerated
electrons on $\hbar\omega$ (Ref.~\onlinecite{Meier18}) and also
due to spin relaxation of electrons during the diffusion towards
the emitting surface.

In order to determine the spin diffusion length and the degree of
spin polarization of photoemitted electrons, we compared the
experimental dependence $A(\hbar\omega)$ with the spectral
dependence of the electron spin polarization calculated in the
diffusion model \cite{Dzhioev26} in the spectral range of
$\varepsilon_g<\hbar\omega<\varepsilon_g+\Delta_{so}$. The
spectral dependence of the initial polarization of photogenerated
electrons and spin depolarization in the process of diffusion
toward the emitting surface were taken into account. The
theoretical dependence is shown in Fig.~6 by the solid curve (the
right scale). The following parameters of the \mbox{AlGaAs/GaAs}
photocathode heterostructure were used for the calculation: the
thickness of the active $p$-GaAs layer, $d=1.2$ $\mu$m; the
electron diffusion length in the active layer, $L_e=3.5$ $\mu$m;
the recombination velocity between the $p$-GaAs and buffer AlGaAs
layers, $V_{S1}=10^4$ cm/s; and the effective recombination
velocity at the emitting surface, $V_{S2}=10^7$ cm/s. The spin
diffusion length $L_S$ was a fitting parameter. As the electron
diffusion length $L_e>d$ and $V_{S1}<<V_{S2}$, the variations of
the parameters $L_e$ (at $L_e>3$ $\mu$m), $V_{S1}$ (from $10^4$ to
$5\times10^5$ cm/s), and $V_{S2}$ (in the range of
$3\times10^6-10^7$ cm/s) do not lead to significant changes in the
calculated curve. The best fit value of the spin diffusion length
was equal to $L_S=0.45\pm0.05$ $\mu$m. The corresponding value of
the spin polarization of photoemitted electrons generated by
photons with energies near the band gap
$\hbar\omega\approx\varepsilon_g$ is equal to $S=0.20\pm0.02$. The
obtained value of the spin diffusion length is in good agreement
with $L_S=0.55$ $\mu$m reported by Dzhioev et al.\cite{Dzhioev26}
The authors of Ref.~\onlinecite{Dzhioev26} studied spin
polarization by the photoluminescence technique in a glass-bonded
photocathode structure similar to that used in our photoemission
experiment, with about the same value of doping level in the
active $p$-GaAs layer.

\section{Conclusions}

Thus, it is experimentally found that in an external magnetic
field the probability of electron emission from GaAs with the
state of NEA into vacuum depends on the orientation of electron
spin with respect to the direction of the field. This phenomenon
stems from the jump in the electron $g$ factor on the
semiconductor-vacuum interface. Due to this jump, the effective
electron affinity of a photocathode depends on the mutual
directions of the electron spin and magnetic field. This mechanism
successfully describes the magnitude and energy dependence of SDP.
A comparison of the measured and calculated spectra of the
spin-dependent photoemission enabled us to determine the spin
diffusion length in $p$-GaAs and spin polarization of photoemitted
electrons.

\begin{acknowledgments}
This work was supported by the Russian Academy of Sciences
(program "Spin-dependent phenomena and spintronics") and by the
Russian Foundation for Basic Research (Grant No. 07-02-01005).
\end{acknowledgments}

\end{document}